\documentclass[pdflatex,sn-mathphys,Numbered]{sn-jnl}

\usepackage{graphicx}%
\usepackage{multirow}%
\usepackage{amsmath,amssymb,amsfonts}%
\usepackage{amsthm}%
\usepackage{mathrsfs}%
\usepackage[title]{appendix}%
\usepackage{xcolor}%
\usepackage{textcomp}%
\usepackage{manyfoot}%
\usepackage{booktabs}%
\usepackage{algorithm}%
\usepackage{algorithmicx}%
\usepackage{algpseudocode}%
\usepackage{listings}%
\usepackage[nameinlink]{cleveref}
\setcitestyle{compress,round,authoryear}
\theoremstyle{thmstyleone}%
\theoremstyle{thmstyletwo}%

\theoremstyle{thmstylethree}%

\begin{document}

\title[BARN]{BARMPy: Bayesian Additive Regression Models Python Package}


\author[1,2]{\fnm{Danielle} \sur{Van Boxel}}\email{vanboxel@math.arizona.edu}



\affil[1]{\orgdiv{Applied Math GIDP}, \orgname{University of Arizona}, \orgaddress{\city{Tucson}, \state{AZ}, \country{USA}}}
\affil[2]{\orgdiv{Data Diversity Lab}, \orgname{University of Arizona}, \orgaddress{\city{Tucson}, \state{AZ}, \country{USA}}}




\abstract{
    We make Bayesian Additive Regression Networks (BARN) available as a Python package, \texttt{barmpy}, with documentation at \url{https://dvbuntu.github.io/barmpy/} for general machine learning practitioners.  Our object-oriented design is compatible with SciKit-Learn, allowing usage of their tools like cross-validation.  To ease learning to use \texttt{barmpy}, we produce a companion tutorial that expands on reference information in the documentation.  Any interested user can \texttt{pip install barmpy} from the official PyPi repository.  \texttt{barmpy} also serves as a baseline Python library for generic Bayesian Additive Regression Models.
}

\keywords{machine learning, Python, MCMC, software}



\maketitle

\section{Introduction}

We implement Bayesian Additive Regression Networks (BARN) as a software package, \texttt{barmpy} (for Bayesian Additive Regression \emph{Models} in Python).  This algorithm is another approach to the general regression problem of finding some function, $f(x_i)$, to approximate a noisy relationship, $y_i = u(x_i) + \epsilon_i$, where $\epsilon_i \sim N(0,\sigma)$ is a noise term and $u(x_i)$ represents some true underlying function.  BARN works by sampling from a posterior distribution on ensembles of neural networks, similar to Bayesian Additive Regression Trees (BART) \citep{chipman2010bart}, but with a different backbone.  We cover some of the necessary practical mathematical background in \Cref{sec:math}, but here we focus more on implementation designs for the library.

New methods in machine learning and broader mathematics arise all the time, but they are not always readily accessible to data science practitioners \citep{patel2008examining}.  Part of the explosion of machine learning was the development of libraries like Scikit-Learn \citep{scikit-learn}, TensorFlow \citep{tensorflow2015whitepaper}, and Keras \citep{chollet2015keras}.  Such tools not only freed data scientists from having to implement machine learning algorithms manually, they also provide detailed documentation with examples.  This kind of broad support is the difference between a research algorithm and an accessible library.

Making \texttt{barmpy} accessible means more than publishing code.  We integrate tightly with existing popular Python machine learning libraries like Scikit-Learn (and TensorFlow as an alternative), described in more detail in \Cref{sec:design}.  This includes following those libraries' best practices regarding complete documentation and example tutorials.  We even take the algorithmic improvement beyond Scikit-Learn by implementing custom model callbacks, such as for early stopping, detailed in \Cref{subsec:early}.  To show \texttt{barmpy}'s utility and limitations, we conduct benchmark testing and a small case study.  Part of this is computation time information, as such metrics are often a concern for machine learning practitioners.  We note that computation time is itself an accessibility issue; large language models like ChatGPT are not generally trainable to users with typical hardware resources \citep{ouyang2022training}.  Making a new method like BARN usable in terms of speed, capability, and understandability makes \texttt{barmpy} more than an algorithm.

\section{Mathematical Background}\label{sec:math}

Whereas we describe the methodology of BARN in detail in \citet{vanboxel2024barn}, here we review key points, as relevant to potential users, of BARN as implemented in \texttt{barmpy}.  Recall that the related method, BART, is made of an ensemble of decision trees which sum to the prediction \citep{chipman2010bart}.  While structurally similar to a random forest \citep{breiman2001random}, BART is distinct in that we train it by sampling from the posterior distribution of trees.  By carefully setting transition, prior, and evidence probability functions, BART can calculate an MCMC acceptance ratio.  This is the chance of accepting a changed tree.  After many iterations over all the trees, we realize convergence to the desired posterior.  BARN works similarly to this but uses neural networks in the ensemble rather than decision trees.  From an algorithmic (and software design) perspective, however, we need to define the MCMC steps of model proposal, transition, and posterior.

Consider the transition probabilities, which in a way, encapsulate the model proposal.  In BARN, as in BART, we apply Gibbs sampling to ensembling, proposing, training, and potentially accepting a single neural network at a time.  BARN allows only 2 transitions: adding one neuron or subtracting one neuron.  Therefore, we capture this in a single parameter, $p$, the probability of adding a neuron to the existing network.  This proposes the new network size, but to fully specify that network, we also need model weights.  That involves training the network with standard optimization techniques, again as described in \citet{vanboxel2024barn}.  The final proposed new network in a step is then the result of this procedure.

To compute the MCMC acceptance ratio, we also need the posterior probability of the old and new networks.  Following Bayes' rule, we use the prior probability times the evidence probability (ignoring the data probability as that will cancel in the ratio).  Note that in BART, this calculation is the closed form of an integral over the weights \citep{chipman2010bart}.  In BARN, this is an approximation, so we only need a closed form for the prior and evidence.  Our default prior depends only on the number of neurons and uses a discrete Poisson distribution.  And finally, the evidence component of the BARN posterior for model $k$ is the likelihood of the target residual value, $P(R_k|M_k,X)$, where $R_k = Y - \sum_{j\neq k} M_j(X)$ and $M_j(X)$ is the $j$th neural network in the ensemble applied to input $X$.   This likelihood assumes a normal distribution of errors (and sampled $\sigma$ value for each MCMC step).  The prior times this evidence gives us the posterior, which then multiplied by the transition proposal probability, contributes to the acceptance ratio.  This provides all the key aspects of a BARN model.

As briefly mentioned, \texttt{barmpy} users can supply their own parameters or methods for these inputs.  \Cref{subsec:cvtune} describes in detail how to use Scikit-Learn's own cross-validation with \texttt{barmpy}.  And because the library is object-oriented, data scientists familiar with Python and Scikit-Learn can quickly subclass our \texttt{BARN} or \texttt{NN} classes to their own specifications.  A good use case for this would adapt BARN for binary classification by changing a few methods that control the MCMC process.  So even if \texttt{barmpy} basic methods and defaults are not directly applicable, they can serve as a starting point for rapid prototyping.

\section{Library Features}\label{sec:design}

In developing \texttt{barmpy}, we seek not only to implement BARN for regression and classification, but also to create an accessible library for generic BART or BARN-like algorithms.  Part of that means weighing different programming language options, not only for ease of our own coding, but for future open-source development as well.  Additionally, we explore some of the choices behind the overall object-oriented design.  This design includes tight integration with Scikit-Learn \citep{scikit-learn}, which multiplies \texttt{barmpy}'s capabilities.  Next, we note the importance of not only documentation, but fully worked tutorials for practitioners.  And finally, we discuss some practical concerns like distribution on PyPi \citep{pypi} and GitHub \citep{escamilla2022rise}.  Our goals in developing \texttt{barmpy} go beyond merely implementing accurate statistics.

\subsection{Design Considerations}

We choose to develop \texttt{barmpy} in Python, even though most BART packages are written in R.  The original researchers into BART are responsible for multiple packages in R, including variants \citep{chipman2022package,mcculloch2024package}, generally available on the Comprehensive R Archive Network (CRAN) \citep{hornik2012comprehensive}.  Other BART-derived methods like MOTR-BART also develop in R \citep{prado2021bayesian}.  While BARN is related to BART, we intend \texttt{barmpy} to be of general use not only to statisticians, but to professional data scientists as well.  And these data scientists almost overwhelmingly choose Python, partly due to its broader ecosystem of libraries, documentation, and developer tools \citep{srinath2017python}.  Therefore, we develop \texttt{barmpy} to reach the data scientists where they are.

One particularly relevant Python library we import is Scikit-Learn \citep{scikit-learn}, which implements a huge variety of machine learning algorithms with a standardized object-oriented application program interface (API).  While libraries like \texttt{statsmodels} provide detailed statistics in a manner similar to R packages \citep{seabold2010statsmodels}, Scikit-Learn focuses on practical usage and extensibility.  For example, \texttt{statsmodels} requires some workarounds to enable using models for prediction on new data, but every Scikit-Learn \texttt{Estimator} has a built-in \texttt{predict} method for exactly this.  Further, Scikit-Learn implements a method for cross-validated hyperparameter tuning; anything that subclasses a Scikit-Learn \texttt{Estimator} may tune this way.  \texttt{barmpy} benefits from many of these Scikit-Learn features.

\subsection{Regression and Binary Classification}

As our BARN ensemble is made of many small neural networks, our fundamental class is \texttt{NN}, which uses the Scikit-Learn primitive, \texttt{MLPRegressor} (for ``Multi-Layer Perceptron'', i.e. a fully connected neural network).  A short example of training BARN is in \Cref{alg:reg_class}.  Our class includes helper methods to compute the various MCMC log-likelihood and prior probabilities given a network, as this is done on a per-network level under Gibbs sampling.  There are also routines to quickly save or load results, and handle the weight donation to other neural networks when transitioning.  Building an ensemble of these \texttt{NN} objects, we have the general \texttt{BARN} class.  This is more than a list of \texttt{NN} objects; it includes parameters to customize the algorithm priors.  Further, it has a critical method, \texttt{BARN.fit}, which implements the full BARN procedure with Gibbs sampling.  While not parallelized (as models must be fit sequentially), it does avoid duplication of computation by caching residual values and only updating them with the networks that have changed.  This turns an $O(N)$ operation, where $N$ is the number of networks in the ensemble, into an $O(1)$ (i.e. constant time) operation.  And like \texttt{NN}, this class has some helper methods for features like Monte Carlo batch means analysis and built-in visualization of key results using \texttt{matplotlib} \citep{Hunter:2007}.  From a user perspective, they need only instantiate a \texttt{BARN} object, setup the Bayesian parameters, and supply data for training.

BARN for classification works similar to regression.  Currently, the library supports binary classification with targets encoded as $y_i \in \{0,1\}$.  In code, one swaps \texttt{BARN\_bin} for \texttt{BARN}.  After training, \texttt{BARN\_bin}'s predictions lie in $(0,1)$, and represent the model's predicted probability of the true class being 1, as in a probit model.  If desired, one can directly produce the model $z$-scores which equate to these probabilities.  For usage, we again offer a small example in \Cref{alg:reg_class}.  Internally, \texttt{BARN\_bin} inherits from the same base class as \texttt{BARN}, \texttt{BARN\_base}, which implements most of the sampling logic.  Because of this, \texttt{BARN\_bin} uses the same \texttt{MLPRegressor} (not \texttt{MLPClassifier}) for each component of the ensemble.  Options like prior distribution mean value are identical, making it easy to switch between these BARN modes for different problems as needed.

\begin{algorithm}
    \caption{BARN in regression and classification using \texttt{sklearn}}\label{alg:reg_class}
\begin{lstlisting}[language=Python]
from sklearn import datasets
import sklearn.metrics
from barmpy.barn import BARN, BARN_bin
import numpy as np

# Regression problem
db = datasets.load_diabetes()
model = BARN(num_nets=10,
          random_state=0,
          warm_start=True,
          solver='lbfgs',
		  l=1)
model.fit(db.data, db.target)
pred = model.predict(db.data)
print(sklearn.metrics.r2_score(db.target, pred))

# Classification problem
bc = datasets.load_breast_cancer()
bmodel = BARN_bin(num_nets=10,
          random_state=0,
          warm_start=True,
          solver='lbfgs',
		  l=1)
bmodel.fit(bc.data, bc.target)
pred = bmodel.predict(bc.data)
print(sklearn.metrics.classification_report(bc.target, np.round(pred)))
\end{lstlisting}
\end{algorithm}

Our BARN implementation comes with reasonable defaults for ease of use by scientific practitioners.  We recommend the NN growth transition probability be set to $p=0.4$.  This mildly encourages the algorithm to test relatively small networks, mitigating the chance of a single network dominating the ensemble.  Similarly, we advise setting the network size prior distribution mean to  $\lambda =1$ or another small value to again encourage networks to be individually weak learners.  If one is using BARN for pure architecture search (i.e. only a single network in the ensemble), however, then $\lambda$ should be larger to accommodate more complexity.  Additionally, users can supply their own generic probability mass function if they wish to control the prior more carefully.  The number of networks in the ensemble, as mentioned, is itself a settable parameter.  We default to 10 as this balances ensembling to improve generalization with increased computation time from additional network training.  For the neural network training itself, parameters like learning rate and weight regularization penalties are more problem-dependent.  We suggest learning rate $lr=0.01$ and L1L2 regularization $r=0.01$, but note that users should experiment with these particular settings.  Additional details on tweakable parameters are available in the BARN documentation \citep{vanboxel2023barmdoc}.

\subsection{Improving Software Accessibility}

\begin{figure}[htb]
\centering
    \includegraphics[scale=0.75]{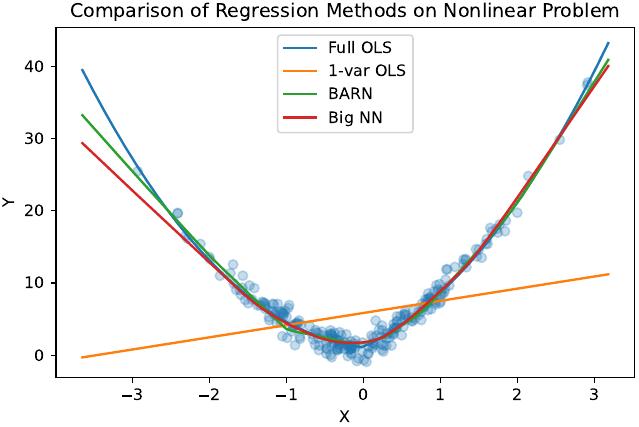}
    \caption{Example tutorial output showing BARN may outperform OLS and even a much larger neural network.  Note this example uses no cross-validated tuning.}
    \label{fig:tut}
\end{figure}

To further ease usage and additional development, we have adopted several more general software engineering principles.  First, being a Python library, we naturally distribute \texttt{barmpy} as a package on PyPi.org \citep{pypi}.  This enables hassle-free installation for new users.  Next, as noted earlier, we maintain all development history in a Git repository on Github.  To ensure functional correctness even in the face of seemingly unrelated changes, we run a suite a unit tests with every commit to the \texttt{main} branch.  Each test runs a small chunk of code using \texttt{barmpy} as a library and compares the output to a known good result.  When a test fails, we can see exactly where and if this needs attention.  In addition to assisting with development, unit test also act as examples for new users.  Beyond such rigorous tests, we also wrote and deployed a complete walkthrough via an R Markdown \citep{baumer2015r} script.  This walkthrough describes a problem end-to-end, from generating data to running BARN and interpreting the results.  Further, because this is in R Markdown, users can run the code chunks themselves (by \texttt{knit}ting the script or copying it into a Python terminal).  Example output is provided, as in \Cref{fig:tut}, for users to verify their results, thereby ensuring they can learn how to apply \texttt{barmpy} on their own.  Finally, when users or developers need more details, they can review the low-level documentation we developed using Python's Sphinx library \citep{brandl2021sphinx}.  This documentation is in part automatically generated from the BARN Python docstrings themselves, though we include additional mathematical information at an appropriate level, such as for doing cross-validation on BARN with Scikit-Learn.  The source documents are part of the repository itself, but they are also online as a Github Pages website \citep{vanboxel2023barmdoc}.   These tools enable new users and developers to quickly understand, use, and improve on the BARN algorithm for data science projects.

\subsection{Cross-Validated Tuning Via Scikit-Learn}\label{subsec:cvtune}

BARN is implemented as a \texttt{sklearn} class, meaning we can use standard \texttt{sklearn} methods like \texttt{GridSearchCV} to tune the hyperparameters for the best possible result.  Note that each additional parameter choice increases the computation time multiplicatively, so one should be mindful when considering the number of possible hyperparameter values.

All arguments to \texttt{BARN} which accept different values can be tuned this way.  In \Cref{alg:cv}, we show an example that tunes the prior distribution mean value parameter, $\lambda$.  Also, when using a method like \texttt{RandomizedSearchCV}, one should be careful to supply appropriate distributions. Here, $l$ takes discrete values, so we specify a discrete Poisson probability distribution to sample from. Note, however, that this distribution is only for cross-validation sampling of the prior parameters, not for BARN to use in internal MCMC transitions.

\begin{algorithm}
\caption{BARN CV Tuning Example using \texttt{sklearn}}\label{alg:cv}
\begin{lstlisting}[language=Python]
from sklearn import datasets
from sklearn.model_selection import GridSearchCV, RandomizedSearchCV
from barmpy.barn import BARN
db = datasets.load_diabetes()
scoring = 'neg_root_mean_squared_error'

# exhaustive grid search
## first make prototype with fixed parameters
bmodel = BARN(num_nets=10,
          random_state=0,
          warm_start=True,
          solver='lbfgs')
## declare parameters to exhaust over
parameters = {'l': (1,2,3)}
barncv = GridSearchCV(bmodel, parameters,
                refit=True, verbose=4,
                scoring=scoring)
barncv.fit(db.data, db.target)
print(barncv.best_params_)

# randomized search with distributions
from sklearn.model_selection import RandomizedSearchCV
from scipy.stats import poisson
## first make prototype with fixed parameters
bmodel = BARN(num_nets=10,
          random_state=0,
          warm_start=True,
          solver='lbfgs')
## declare parameters and distributions
parameters = {'l': poisson(mu=2)}
barncv = RandomizedSearchCV(bmodel, parameters,
                refit=True, verbose=4,
                scoring=scoring, n_iter=3)
barncv.fit(db.data, db.target)
print(barncv.best_params_)
\end{lstlisting}
\end{algorithm}

\subsection{Early Stopping Approaches}\label{subsec:early}

In machine learning, even when training some model over many iterations, it is common to stop the process early under some conditions.  Typically, these involve checking some error metric against held-out validation data \citep{genccay2001pricing}.  If the metric fails to improve, then one stops training in order to avoid overfitting to training data.  Given the MCMC-based training process of BARN, however, there are several possibilities for such metrics.

In addition to the standard approach of checking validation error, we explore alternatives measuring stability in the posterior.  As the MCMC posterior is some probability distribution, we can estimate it from our samples once we reach convergence.  If this estimate is stable, then we infer that convergence has been reached and we can stop.  One reasonable metric is the earth-mover distance (also known as the one-Wasserstein metric \citep{solomon2014earth}) from one estimate of the distribution to the next.  In our case, this means evaluating the distribution of the number of neurons in each network of the ensemble and setting a change threshold.  Though some researchers explored similar ideas \citep{durmus2015quantitative}, they focused more on mixing rate.  A similar though simpler heuristic is to simply check how many proposed model transitions BARN accepted in the previous iteration.  If the model has converged, then the error rates are already low and it will be relatively difficult to dislodge an existing model.  Therefore, so long as a model continues to accept transitions, it advances to the next MCMC iteration, as \Cref{alg:trans} details in Python.  By default, this method stops if less than 20\% of the networks in the ensemble transitioned.  A final, more rigorous alternative is to check not just the stationarity of the MCMC calculation, but the complete convergence of batch means as well.  The Relative Fixed-Width Stopping Rule constructs a $t$-stat to check recent convergence of relative batch means, implying stationarity \citep{flegal2015relative}.  These are all relatively quick to implement, so we make them available to users as a model callback.

\begin{algorithm}
\caption{``Not Trans Enough'' Early Stopping Callback}\label{alg:trans}
\begin{lstlisting}[language=Python]
@staticmethod
def trans_enough(self, check_every=None, skip_first=0, ntrans=None):
    '''
    Stop early if fewer than `ntrans` transitions

    Skip the first `skip_first` iters without checking
    '''
    i = self.i
    # only check every so many, default every 10%
    if check_every is None:
        check_every = max(self.n_iter//10, 1)
    # not an iteration to stop on
    if i == 0 or i % check_every != 0:
        return None
    if i < skip_first:
        return None
    # default minimum transitions to continue is 20%
    if ntrans is None:
        ntrans = max(self.num_nets//5,1)
    # compare most recent count of accepted transitions
    if self.ntrans_iter[i-1] < ntrans:
        raise JackPot # early stopping flag
\end{lstlisting}
\end{algorithm}

In practice, however, we expect most data scientists to use the more common check on the current model validation error than these other methods.  In various evaluations, we found most methods provide similar results (about a 20\% reduction in computation time), with validation error anecdotally being the most stable.  We still expose them, not only for their nominal purpose, but also as examples of generic custom model callbacks that can affect the training procedure. 

\section{Evaluation}\label{sec:eval_code}

To see in what contexts \texttt{barmpy} is most useful, we analyze its error and timing metrics in different situations.  We focus on both a small case study with data from an active problem in biology as well as a review of computation time on different synthetic data sets.

Before discussing these results, we quickly note how \citet{vanboxel2024barn} covers a broad range of real and synthetic data sets to show where BARN is most effective.  In particular, their analysis of specific synthetic data sets provides some of the most insight.  Without repeating the analysis there, we note that BARN does better than other methods on problems where there is a strong functional nonlinear relationship like Friedman $F2$ or $F3$.  So BARN may be practically appropriate as an approximation to a complex system that cannot be easily directly modeled.

\subsection{Case Study: Isotope Modeling}\label{subsec:isotope}

While \citet{vanboxel2024barn} runs BARN on a wide variety data sets, we focus here on one case study on clumped isotope paleothermometry.  The modeling problem itself is to predict carbonate clumped isotope thermometry, $\Delta_{47}$, as a function of temperature \citep{eiler200418o13c16o}.  This is a calibration process; in practice one uses $\Delta_{47}$ as a surrogate for historical temperature that was not measured (and therefore something invertible like OLS is typically preferable).  The ecological details are beyond the scope of this paper, but there are various studies on this topic \citep{eiler200418o13c16o,roman2022bayclump,petersen2019effects}.  A recent study \citep{roman2022bayclump} demonstrated the effectiveness of a Bayesian least squares approach to this data.  Such a method uses a linear model as in OLS, but employs priors on the estimated parameter values informed by earlier studies.  BARN also uses priors but on the model structure (by affecting the size of learned networks) rather than the parameters directly.

\begin{figure}[ht]
\centering
    \includegraphics[scale=.7]{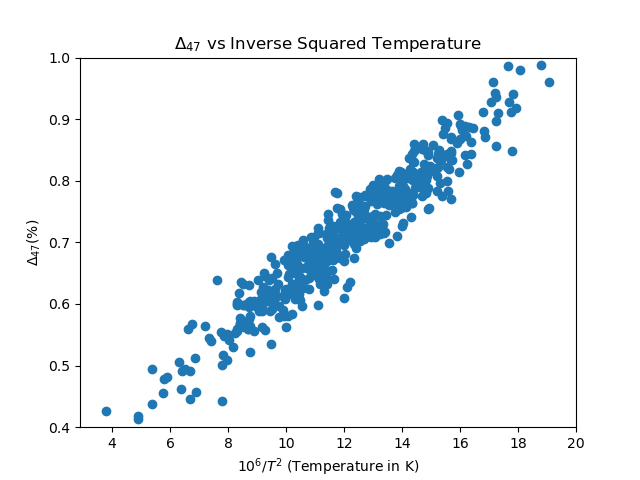} 
    \caption{Apparent linear relationship between temperature and $\Delta_{47}$}
    \label{fig:d47}
\end{figure}

As this data set is in a single variable, we can visually inspect the relationship between temperature and $\Delta_{47}$.  \Cref{fig:d47} plots $\Delta_{47}$ against the inverse of squared temperature, showing a strong linear relationship, though with some spread.  Scientists training models on this data need to be able to invert the model (i.e. change $\Delta_{47} = f(T)$ into $T = f^{-1}(\Delta_{47})$) to predict historical temperatures.  So even if BARN outperforms other approaches, it will likely not replace linear methods on this particular problem.  We focus on BARN's performance on the data as an area of active research.

\begin{figure}[ht]
\centering
    \includegraphics[scale=.7]{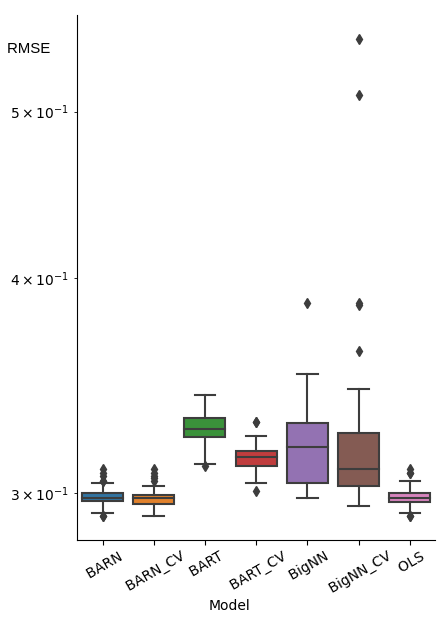}
    \caption{Absolute RMSE boxplot of various methods on the isotope data set.  BARN (with or without tuning) and OLS have similar profiles, while other methods are significantly worse (but still rather accurate; note log scale).}
    \label{fig:isotope}
\end{figure}

In \Cref{fig:isotope} we inspect results on this ``isotope'' data set, and we see that BARN performs well relative to the other methods.  Note that the output has been rescaled from the original for this calculation.  BARN produces very similar results to OLS (test RMSE about 0.298, 4\% less than the next-best method's error).  This performance does require a greatly increased computational cost, as we shall see in \Cref{subsec:cost}.  We caution, again, that our BARN analysis here is for demonstration only, as this particular problem requires invertibility.

\begin{figure}[ht]
\centering
    \includegraphics[scale=.7]{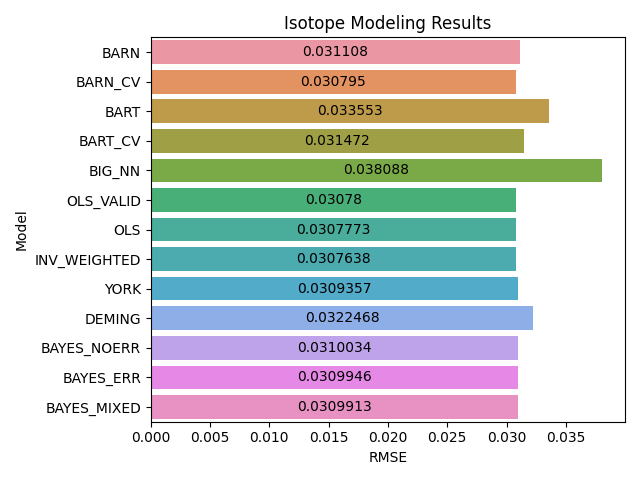}
    \caption{Testing RMSE point estimates (only a single split performed) on $\Delta_{47}$ testing data for various methods.  BARN performs similarly to linear methods \citep{roman2022bayclump} even when a big NN and BART perform significantly worse.  Note that methods used in this study (top six models) reserve 25\% of the training data for validation (hence why ``OLS\_VALID'' is separate from ``OLS'').}
    \label{fig:iso_results}
\end{figure}

The state-of-the art in this area uses Bayesian linear models.  We show \Cref{fig:iso_results} to quickly compare existing methods with our approaches on a specific data split of interest (hence why there are only point estimates of the error).  Further, we note that these are on the \emph{original data scales}, hence why all the errors appear so much smaller than in \Cref{fig:isotope}.  BARN appears to be in the same class of error levels as the best linear approaches.  BARN's error is only about 1\% higher than the best method (and even closer for tuned BARN).  This is especially interesting because the other nonlinear methods we tested (the big NN and BART) actually perform significantly \emph{worse} than OLS and BARN when not using cross-validated tuning.  It is possible that BARN is able to simplify to an OLS-like model that is appropriate for this problem (which has a single explanatory variable) where other nonlinear methods would require additional training data for such a reduction.  This demonstrates some of the adaptability and broad applicability of BARN.

\subsection{Computational Costs of BARN}\label{subsec:cost}

While \citet{vanboxel2024barn} described BARN's performance in terms of error, here we consider the computational cost of running it.  There can be a trade-off here.  Some problems, like targeted display advertising \citep{shah2020research}, benefit from speed of computation (at the expense of accuracy); others, like medical imaging \citep{aggarwal2021diagnostic}, require very low prediction error and are willing to invest computational resources to achieve it.  To assess BARN on this axis, we consider the data sets described in \citet{vanboxel2024barn}.  While these data sets are modest in size (about 1000 data points and 10 features), they are sufficient to realize the differences in computation times, in seconds, shown in \Cref{tab:times}.  These times do vary across runs, but not to the extent of the order of magnitude differences in times across methods.

\begin{figure}[htb]
\centering
    \includegraphics[scale=.6]{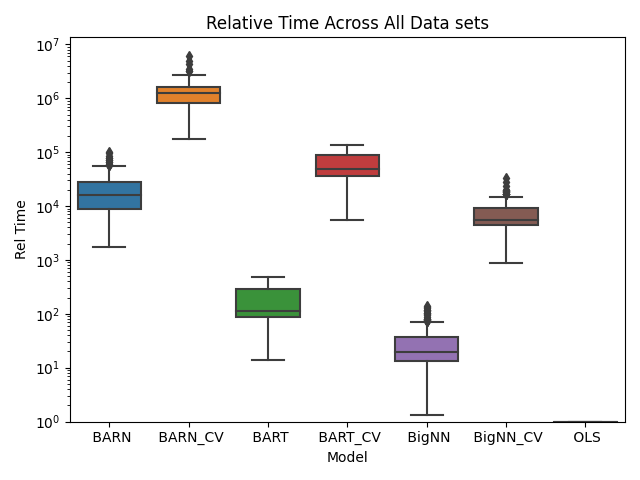}
    \caption{BARN is comparable in time to other methods with cross-validation}
    \label{fig:times}
\end{figure}

In \Cref{fig:times}, we see the relative computation times for our various methods on all data sets.  OLS, being only linear algebra, is always the fastest (hence the relative time of 1).  Next, training a single neural network with gradient descent takes 10 to 100 times as long (still less than a second on any given problem).  BART is solidly 100 times slower than OLS, about 1 second of real time.  Plain BARN is about 10,000 times slower than OLS, taking on the order of 10 seconds to a 1 minute on a given problem.  This is primarily due to the necessity of training new small neural networks for all the MCMC iterations.  While each one is very fast (close to 10 times faster than the single big network), doing this for 200 MCMC iterations is a significant cost.  BART avoids this cost because it does not ``train'' in a traditional sense (i.e. it does not set weights with the standard CART procedure), so the MCMC iterations are not as computationally intense.  On the face of it, BARN seems like it is very slow.

When we consider the methods with cross-validated hyperparameter tuning, however, we see that BARN is actually time-competitive with the other nonlinear methods.  Those methods are on the order of tens of seconds for this data.  Now, from earlier studies \citep{vanboxel2024barn}, we know that plain BARN is nearly as accurate as BARN with tuning.  Yet plain BARN still produces lower error than BART or the big neural network \emph{even when those methods are tuned}.  Looking at \Cref{fig:times} again,  we see that plain BARN takes about the same amount of relative time as other methods when those are tuned.  Those methods benefit significantly from such tuning, whereas BARN may be adaptable even without it.  For situations requiring low testing error in regression, BARN is time-competitive with other nonlinear algorithms.

\begin{table}[htb]
\centering
\caption{Mean training time in seconds over 40 trials of various methods on different data sets}
\begin{tabular}{lrrrrrrr}
    Dataset   &     BARN &  BARN CV & BART & BART CV & Big NN &  Big NN CV & OLS\\ \hline
cali small &  70.315  &  3051.940 & 0.821  &   240.575  & 0.149 &    29.332 & 0.002  \\  
concrete   &  76.898  &  3418.810 & 0.494  &   148.01   & 0.066 &    20.202 & 0.002    \\
crimes     & 270.220  & 14175.800 & 0.905  &   258.248  & 0.183 &    30.963 & 0.011    \\
diabetes   &  16.364  &  2231.910 & 0.177  &    73.3536 & 0.035 &     9.663 & 0.002    \\
fires      &  33.866  &  3302.060 & 0.168  &    78.0964 & 0.036 &    10.802 & 0.002    \\
isotope    &   7.521  &   526.695 & 0.350  &   136.935  & 0.032 &     6.532 & 0.001    \\
mpg        &  30.834  &  1923.640 & 0.170  &    66.4122 & 0.031 &     8.533 & 0.002    \\
random     &  21.660  &  1117.350 & 0.574  &   146.616  & 0.039 &    15.839 & 0.002  \\
wisconsin  &  58.132  &  3734.430 & 0.120  &    48.0919 & 0.043 &     7.964 & 0.002    \\
\end{tabular}
    \label{tab:times}
\end{table}

We did explore various speedups to BARN.  Recall that we chose to implement BARN in a Scikit-Learn compatible way, including using their \texttt{MLPRegressor} class.  While this is convenient, it may not be the fastest NN implementation.  So we also implemented BARN using TensorFlow-based neural networks, including training these NNs on GPUs \citep{tensorflow2015whitepaper}.  For large neural networks, GPUs often provide a 5 times or better speedup from parallelism \citep{lind2019performance}, but with BARN, we found just the opposite.  Because BARN typically uses tiny neural networks, these map poorly to GPUs.  Significant time, relative to the computation cost, is lost simply moving data between CPU and GPU.  Another improvement we tried, however, was more helpful.  Initially, we trained BARN networks using typical gradient-descent style solvers common in machine learning and TensorFlow.  But since we were using Scikit-Learn, we also had easily available quasi-Newton methods like BFGS \citep{nocedal1999numerical}.  At the scale of networks and data sets considered, we found switching away from gradient descent provided a 2 to 4 times speedup.  Understanding that this is problem dependent, however, we enable setting this parameter in the function and provide a sane default that selects based on network size.  Such techniques provide BARN with some speed enhancements, though more research in this area is needed.

\section{Conclusion}\label{sec:soft_conc}

We reviewed the design and capabilities of the new Python package, Bayesian Additive Regression Models in Python (\texttt{barmpy}).  In addition to the favorable results of lower error rates on benchmark data seen in \citet{vanboxel2024barn}, we found \texttt{barmpy} to be fast enough on relevant problems.  While it is an order of magnitude slower than BART, ARN does not need hyperparameter tuning to do well, making it generally time-competitive.  Still, additional research into faster implementation of BARN would be beneficial.  TensorFlow wasn't able to improve speeds, but another linear algebra library, one tuned for many small matrices, might be appropriate.  Or, BARN may benefit from an algorithmic change.  For example, rather than learning weights via the neural network training procedure, we could sample them directly as part of the MCMC process.  This has the downside of ignoring existing optimization approaches, but something similar works for BART, so it may work here.  Beyond direct metrics, we also emphasized the importance of accessibility for \texttt{barmpy}.  This is why we chose to develop it in Python with tight integration with Scikit-Learn.  We meet the practitioners where they are.  Likewise, we recognize the importance of self-teaching in learning new software.  So we provide not only the library itself, but supporting documentation and tutorials.  Finally, we consider some future additions to the package.  As the name, \texttt{barmpy}, suggests, we seek to support a generic model backbone, not just neural networks.  Provided one can (hopefully rigorously) supply transition and posterior probability methods, this ought to be broadly applicable.  For example, support vector machines may be a straightforward next backbone option to implement.  All these implementation details and other ``extras'' are necessary for enabling users to learn \texttt{barmpy} and employ it effectively.

\backmatter

\bmhead{Acknowledgments}

I must thank both of my PhD co-advisors, Xueying Tang and Cristian Rom\'an-Palacios, for their constant guidance and support.  Prof. Tang provided key mathematical insight and ensured ongoing statistical rigor.  Prof. Rom\'an-Palacios balanced this with practical machine learning advice as well as the perspective of a research scientist.

\section*{Declarations}

\subsection*{Competing Interests}

This research was performed in part while employed by the Data Diversity Lab within the School of Information at The University of Arizona.

\subsection*{Code Availability}

The \texttt{barmpy} library is available in full on GitHub at \url{https://github.com/dvbuntu/barmpy}.  

\bibliography{lib.bib}


\begin{thebibliography}{31}
\ifx \bisbn   \undefined \def \bisbn  #1{ISBN #1}\fi
\ifx \binits  \undefined \def \binits#1{#1}\fi
\ifx \bauthor  \undefined \def \bauthor#1{#1}\fi
\ifx \batitle  \undefined \def \batitle#1{#1}\fi
\ifx \bjtitle  \undefined \def \bjtitle#1{#1}\fi
\ifx \bvolume  \undefined \def \bvolume#1{\textbf{#1}}\fi
\ifx \byear  \undefined \def \byear#1{#1}\fi
\ifx \bissue  \undefined \def \bissue#1{#1}\fi
\ifx \bfpage  \undefined \def \bfpage#1{#1}\fi
\ifx \blpage  \undefined \def \blpage #1{#1}\fi
\ifx \burl  \undefined \def \burl#1{\textsf{#1}}\fi
\ifx \doiurl  \undefined \def \doiurl#1{\url{https://doi.org/#1}}\fi
\ifx \betal  \undefined \def \betal{\textit{et al.}}\fi
\ifx \binstitute  \undefined \def \binstitute#1{#1}\fi
\ifx \binstitutionaled  \undefined \def \binstitutionaled#1{#1}\fi
\ifx \bctitle  \undefined \def \bctitle#1{#1}\fi
\ifx \beditor  \undefined \def \beditor#1{#1}\fi
\ifx \bpublisher  \undefined \def \bpublisher#1{#1}\fi
\ifx \bbtitle  \undefined \def \bbtitle#1{#1}\fi
\ifx \bedition  \undefined \def \bedition#1{#1}\fi
\ifx \bseriesno  \undefined \def \bseriesno#1{#1}\fi
\ifx \blocation  \undefined \def \blocation#1{#1}\fi
\ifx \bsertitle  \undefined \def \bsertitle#1{#1}\fi
\ifx \bsnm \undefined \def \bsnm#1{#1}\fi
\ifx \bsuffix \undefined \def \bsuffix#1{#1}\fi
\ifx \bparticle \undefined \def \bparticle#1{#1}\fi
\ifx \barticle \undefined \def \barticle#1{#1}\fi
\bibcommenthead
\ifx \bconfdate \undefined \def \bconfdate #1{#1}\fi
\ifx \botherref \undefined \def \botherref #1{#1}\fi
\ifx \url \undefined \def \url#1{\textsf{#1}}\fi
\ifx \bchapter \undefined \def \bchapter#1{#1}\fi
\ifx \bbook \undefined \def \bbook#1{#1}\fi
\ifx \bcomment \undefined \def \bcomment#1{#1}\fi
\ifx \oauthor \undefined \def \oauthor#1{#1}\fi
\ifx \citeauthoryear \undefined \def \citeauthoryear#1{#1}\fi
\ifx \endbibitem  \undefined \def \endbibitem {}\fi
\ifx \bconflocation  \undefined \def \bconflocation#1{#1}\fi
\ifx \arxivurl  \undefined \def \arxivurl#1{\textsf{#1}}\fi
\csname PreBibitemsHook\endcsname

\bibitem[\protect\citeauthoryear{Chipman et~al.}{2010}]{chipman2010bart}
\begin{barticle}
\bauthor{\bsnm{Chipman}, \binits{H.A.}},
\bauthor{\bsnm{George}, \binits{E.I.}},
\bauthor{\bsnm{McCulloch}, \binits{R.E.}}, \betal:
\batitle{{BART}: Bayesian additive regression trees}.
\bjtitle{The Annals of Applied Statistics}
\bvolume{4}(\bissue{1}),
\bfpage{266}--\blpage{298}
(\byear{2010})
\end{barticle}
\endbibitem

\bibitem[\protect\citeauthoryear{Patel et~al.}{2008}]{patel2008examining}
\begin{bchapter}
\bauthor{\bsnm{Patel}, \binits{K.}},
\bauthor{\bsnm{Fogarty}, \binits{J.}},
\bauthor{\bsnm{Landay}, \binits{J.A.}},
\bauthor{\bsnm{Harrison}, \binits{B.L.}}:
\bctitle{Examining difficulties software developers encounter in the adoption
  of statistical machine learning.}
In: \bbtitle{AAAI},
pp. \bfpage{1563}--\blpage{1566}
(\byear{2008})
\end{bchapter}
\endbibitem

\bibitem[\protect\citeauthoryear{Pedregosa et~al.}{2011}]{scikit-learn}
\begin{barticle}
\bauthor{\bsnm{Pedregosa}, \binits{F.}},
\bauthor{\bsnm{Varoquaux}, \binits{G.}},
\bauthor{\bsnm{Gramfort}, \binits{A.}},
\bauthor{\bsnm{Michel}, \binits{V.}},
\bauthor{\bsnm{Thirion}, \binits{B.}},
\bauthor{\bsnm{Grisel}, \binits{O.}},
\bauthor{\bsnm{Blondel}, \binits{M.}},
\bauthor{\bsnm{Prettenhofer}, \binits{P.}},
\bauthor{\bsnm{Weiss}, \binits{R.}},
\bauthor{\bsnm{Dubourg}, \binits{V.}},
\bauthor{\bsnm{Vanderplas}, \binits{J.}},
\bauthor{\bsnm{Passos}, \binits{A.}},
\bauthor{\bsnm{Cournapeau}, \binits{D.}},
\bauthor{\bsnm{Brucher}, \binits{M.}},
\bauthor{\bsnm{Perrot}, \binits{M.}},
\bauthor{\bsnm{Duchesnay}, \binits{E.}}:
\batitle{Scikit-learn: Machine learning in {P}ython}.
\bjtitle{Journal of Machine Learning Research}
\bvolume{12},
\bfpage{2825}--\blpage{2830}
(\byear{2011})
\end{barticle}
\endbibitem

\bibitem[\protect\citeauthoryear{Abadi et~al.}{2015}]{tensorflow2015whitepaper}
\begin{botherref}
\oauthor{\bsnm{Abadi}, \binits{M.}},
\oauthor{\bsnm{Agarwal}, \binits{A.}},
\oauthor{\bsnm{Barham}, \binits{P.}},
\oauthor{\bsnm{Brevdo}, \binits{E.}},
\oauthor{\bsnm{Chen}, \binits{Z.}},
\oauthor{\bsnm{Citro}, \binits{C.}},
\oauthor{\bsnm{Corrado}, \binits{G.S.}},
\oauthor{\bsnm{Davis}, \binits{A.}},
\oauthor{\bsnm{Dean}, \binits{J.}},
\oauthor{\bsnm{Devin}, \binits{M.}},
\oauthor{\bsnm{Ghemawat}, \binits{S.}},
\oauthor{\bsnm{Goodfellow}, \binits{I.}},
\oauthor{\bsnm{Harp}, \binits{A.}},
\oauthor{\bsnm{Irving}, \binits{G.}},
\oauthor{\bsnm{Isard}, \binits{M.}},
\oauthor{\bsnm{Jia}, \binits{Y.}},
\oauthor{\bsnm{Jozefowicz}, \binits{R.}},
\oauthor{\bsnm{Kaiser}, \binits{L.}},
\oauthor{\bsnm{Kudlur}, \binits{M.}},
\oauthor{\bsnm{Levenberg}, \binits{J.}},
\oauthor{\bsnm{Man\'{e}}, \binits{D.}},
\oauthor{\bsnm{Monga}, \binits{R.}},
\oauthor{\bsnm{Moore}, \binits{S.}},
\oauthor{\bsnm{Murray}, \binits{D.}},
\oauthor{\bsnm{Olah}, \binits{C.}},
\oauthor{\bsnm{Schuster}, \binits{M.}},
\oauthor{\bsnm{Shlens}, \binits{J.}},
\oauthor{\bsnm{Steiner}, \binits{B.}},
\oauthor{\bsnm{Sutskever}, \binits{I.}},
\oauthor{\bsnm{Talwar}, \binits{K.}},
\oauthor{\bsnm{Tucker}, \binits{P.}},
\oauthor{\bsnm{Vanhoucke}, \binits{V.}},
\oauthor{\bsnm{Vasudevan}, \binits{V.}},
\oauthor{\bsnm{Vi\'{e}gas}, \binits{F.}},
\oauthor{\bsnm{Vinyals}, \binits{O.}},
\oauthor{\bsnm{Warden}, \binits{P.}},
\oauthor{\bsnm{Wattenberg}, \binits{M.}},
\oauthor{\bsnm{Wicke}, \binits{M.}},
\oauthor{\bsnm{Yu}, \binits{Y.}},
\oauthor{\bsnm{Zheng}, \binits{X.}}:
{TensorFlow}: Large-Scale Machine Learning On Heterogeneous Systems
(2015).
\url{https://www.tensorflow.org/}
\end{botherref}
\endbibitem

\bibitem[\protect\citeauthoryear{Chollet et~al.}{2015}]{chollet2015keras}
\begin{botherref}
\oauthor{\bsnm{Chollet}, \binits{F.}}, et al.:
Keras.
\url{https://keras.io}
(2015)
\end{botherref}
\endbibitem

\bibitem[\protect\citeauthoryear{Ouyang et~al.}{2022}]{ouyang2022training}
\begin{barticle}
\bauthor{\bsnm{Ouyang}, \binits{L.}},
\bauthor{\bsnm{Wu}, \binits{J.}},
\bauthor{\bsnm{Jiang}, \binits{X.}},
\bauthor{\bsnm{Almeida}, \binits{D.}},
\bauthor{\bsnm{Wainwright}, \binits{C.}},
\bauthor{\bsnm{Mishkin}, \binits{P.}},
\bauthor{\bsnm{Zhang}, \binits{C.}},
\bauthor{\bsnm{Agarwal}, \binits{S.}},
\bauthor{\bsnm{Slama}, \binits{K.}},
\bauthor{\bsnm{Ray}, \binits{A.}}, \betal:
\batitle{Training language models to follow instructions with human feedback}.
\bjtitle{Advances in Neural Information Processing Systems}
\bvolume{35},
\bfpage{27730}--\blpage{27744}
(\byear{2022})
\end{barticle}
\endbibitem

\bibitem[\protect\citeauthoryear{}{}]{vanboxel2024barn}
\begin{botherref}
Bayesian additive regression networks
\end{botherref}
\endbibitem

\bibitem[\protect\citeauthoryear{Breiman}{2001}]{breiman2001random}
\begin{barticle}
\bauthor{\bsnm{Breiman}, \binits{L.}}:
\batitle{Random forests}.
\bjtitle{Machine learning}
\bvolume{45}(\bissue{1}),
\bfpage{5}--\blpage{32}
(\byear{2001})
\end{barticle}
\endbibitem

\bibitem[\protect\citeauthoryear{{PyPi Maintainers}}{2023}]{pypi}
\begin{botherref}
\oauthor{\bsnm{{PyPi Maintainers}}}:
Python Package Index - {PyPi}.
Python Software Foundation
(2023)
\end{botherref}
\endbibitem

\bibitem[\protect\citeauthoryear{Escamilla et~al.}{2022}]{escamilla2022rise}
\begin{bchapter}
\bauthor{\bsnm{Escamilla}, \binits{E.}},
\bauthor{\bsnm{Klein}, \binits{M.}},
\bauthor{\bsnm{Cooper}, \binits{T.}},
\bauthor{\bsnm{Rampin}, \binits{V.}},
\bauthor{\bsnm{Weigle}, \binits{M.C.}},
\bauthor{\bsnm{Nelson}, \binits{M.L.}}:
\bctitle{The rise of github in scholarly publications}.
In: \bbtitle{International Conference on Theory and Practice of Digital
  Libraries},
pp. \bfpage{187}--\blpage{200}
(\byear{2022}).
\bcomment{Springer}
\end{bchapter}
\endbibitem

\bibitem[\protect\citeauthoryear{Chipman et~al.}{2022}]{chipman2022package}
\begin{botherref}
\oauthor{\bsnm{Chipman}, \binits{H.}},
\oauthor{\bsnm{McCulloch}, \binits{R.}},
\oauthor{\bsnm{Chipman}, \binits{G.}}:
Package "bayestree"
(2022).
R package version 1.4
\end{botherref}
\endbibitem

\bibitem[\protect\citeauthoryear{McCulloch et~al.}{2024}]{mcculloch2024package}
\begin{botherref}
\oauthor{\bsnm{McCulloch}, \binits{R.}},
\oauthor{\bsnm{Sparapani}, \binits{R.}},
\oauthor{\bsnm{Gramacy}, \binits{R.}},
\oauthor{\bsnm{Pratola}, \binits{M.}},
\oauthor{\bsnm{Spanbauer}, \binits{C.}},
\oauthor{\bsnm{Plummer}, \binits{M.}},
\oauthor{\bsnm{Best}, \binits{N.}},
\oauthor{\bsnm{Cowles}, \binits{K.} \bsuffix{Kate~andVines}}:
Package "bart"
(2024).
R package version 2.9.6
\end{botherref}
\endbibitem

\bibitem[\protect\citeauthoryear{Hornik}{2012}]{hornik2012comprehensive}
\begin{barticle}
\bauthor{\bsnm{Hornik}, \binits{K.}}:
\batitle{The comprehensive r archive network}.
\bjtitle{Wiley interdisciplinary reviews: Computational statistics}
\bvolume{4}(\bissue{4}),
\bfpage{394}--\blpage{398}
(\byear{2012})
\end{barticle}
\endbibitem

\bibitem[\protect\citeauthoryear{Prado et~al.}{2021}]{prado2021bayesian}
\begin{barticle}
\bauthor{\bsnm{Prado}, \binits{E.B.}},
\bauthor{\bsnm{Moral}, \binits{R.A.}},
\bauthor{\bsnm{Parnell}, \binits{A.C.}}:
\batitle{Bayesian additive regression trees with model trees}.
\bjtitle{Statistics and Computing}
\bvolume{31},
\bfpage{1}--\blpage{13}
(\byear{2021})
\end{barticle}
\endbibitem

\bibitem[\protect\citeauthoryear{Srinath}{2017}]{srinath2017python}
\begin{barticle}
\bauthor{\bsnm{Srinath}, \binits{K.}}:
\batitle{Python--the fastest growing programming language}.
\bjtitle{International Research Journal of Engineering and Technology}
\bvolume{4}(\bissue{12}),
\bfpage{354}--\blpage{357}
(\byear{2017})
\end{barticle}
\endbibitem

\bibitem[\protect\citeauthoryear{Seabold and
  Perktold}{2010}]{seabold2010statsmodels}
\begin{bchapter}
\bauthor{\bsnm{Seabold}, \binits{S.}},
\bauthor{\bsnm{Perktold}, \binits{J.}}:
\bctitle{Statsmodels: Econometric and statistical modeling with python}.
In: \bbtitle{Proceedings of the 9th Python in Science Conference},
vol. \bseriesno{57},
pp. \bfpage{10}--\blpage{25080}
(\byear{2010}).
\bcomment{Austin, TX}
\end{bchapter}
\endbibitem

\bibitem[\protect\citeauthoryear{Hunter}{2007}]{Hunter:2007}
\begin{barticle}
\bauthor{\bsnm{Hunter}, \binits{J.D.}}:
\batitle{Matplotlib: A 2d graphics environment}.
\bjtitle{Computing in Science \& Engineering}
\bvolume{9}(\bissue{3}),
\bfpage{90}--\blpage{95}
(\byear{2007})
\doiurl{10.1109/MCSE.2007.55}
\end{barticle}
\endbibitem

\bibitem[\protect\citeauthoryear{Van~Boxel}{2023}]{vanboxel2023barmdoc}
\begin{botherref}
\oauthor{\bsnm{Van~Boxel}, \binits{D.}}:
{barmpy} Documentation.
GitHub
(2023)
\end{botherref}
\endbibitem

\bibitem[\protect\citeauthoryear{Baumer and Udwin}{2015}]{baumer2015r}
\begin{barticle}
\bauthor{\bsnm{Baumer}, \binits{B.}},
\bauthor{\bsnm{Udwin}, \binits{D.}}:
\batitle{R markdown}.
\bjtitle{Wiley Interdisciplinary Reviews: Computational Statistics}
\bvolume{7}(\bissue{3}),
\bfpage{167}--\blpage{177}
(\byear{2015})
\end{barticle}
\endbibitem

\bibitem[\protect\citeauthoryear{Brandl}{2021}]{brandl2021sphinx}
\begin{botherref}
\oauthor{\bsnm{Brandl}, \binits{G.}}:
Sphinx documentation.
\url{http://sphinx-doc.org/sphinx.pdf}
(2021)
\end{botherref}
\endbibitem

\bibitem[\protect\citeauthoryear{Gen{\c{c}}ay and
  Qi}{2001}]{genccay2001pricing}
\begin{barticle}
\bauthor{\bsnm{Gen{\c{c}}ay}, \binits{R.}},
\bauthor{\bsnm{Qi}, \binits{M.}}:
\batitle{Pricing and hedging derivative securities with neural networks:
  Bayesian regularization, early stopping, and bagging}.
\bjtitle{IEEE transactions on neural networks}
\bvolume{12}(\bissue{4}),
\bfpage{726}--\blpage{734}
(\byear{2001})
\end{barticle}
\endbibitem

\bibitem[\protect\citeauthoryear{Solomon et~al.}{2014}]{solomon2014earth}
\begin{barticle}
\bauthor{\bsnm{Solomon}, \binits{J.}},
\bauthor{\bsnm{Rustamov}, \binits{R.}},
\bauthor{\bsnm{Guibas}, \binits{L.}},
\bauthor{\bsnm{Butscher}, \binits{A.}}:
\batitle{Earth mover's distances on discrete surfaces}.
\bjtitle{ACM Transactions on Graphics (ToG)}
\bvolume{33}(\bissue{4}),
\bfpage{1}--\blpage{12}
(\byear{2014})
\end{barticle}
\endbibitem

\bibitem[\protect\citeauthoryear{Durmus and
  Moulines}{2015}]{durmus2015quantitative}
\begin{barticle}
\bauthor{\bsnm{Durmus}, \binits{A.}},
\bauthor{\bsnm{Moulines}, \binits{{\'E}.}}:
\batitle{Quantitative bounds of convergence for geometrically ergodic markov
  chain in the wasserstein distance with application to the metropolis adjusted
  langevin algorithm}.
\bjtitle{Statistics and Computing}
\bvolume{25},
\bfpage{5}--\blpage{19}
(\byear{2015})
\end{barticle}
\endbibitem

\bibitem[\protect\citeauthoryear{Flegal and Gong}{2015}]{flegal2015relative}
\begin{botherref}
\oauthor{\bsnm{Flegal}, \binits{J.M.}},
\oauthor{\bsnm{Gong}, \binits{L.}}:
Relative fixed-width stopping rules for markov chain monte carlo simulations.
Statistica Sinica,
655--675
(2015)
\end{botherref}
\endbibitem

\bibitem[\protect\citeauthoryear{Eiler and Schauble}{2004}]{eiler200418o13c16o}
\begin{barticle}
\bauthor{\bsnm{Eiler}, \binits{J.M.}},
\bauthor{\bsnm{Schauble}, \binits{E.}}:
\batitle{18o13c16o in earth’s atmosphere}.
\bjtitle{Geochimica et Cosmochimica Acta}
\bvolume{68}(\bissue{23}),
\bfpage{4767}--\blpage{4777}
(\byear{2004})
\end{barticle}
\endbibitem

\bibitem[\protect\citeauthoryear{Rom{\'a}n~Palacios
  et~al.}{2022}]{roman2022bayclump}
\begin{botherref}
\oauthor{\bsnm{Rom{\'a}n~Palacios}, \binits{C.}},
\oauthor{\bsnm{Carroll}, \binits{H.}},
\oauthor{\bsnm{Arnold}, \binits{A.}},
\oauthor{\bsnm{Flores}, \binits{R.}},
\oauthor{\bsnm{Petersen}, \binits{S.}},
\oauthor{\bsnm{McKinnon}, \binits{K.}},
\oauthor{\bsnm{Tripati}, \binits{A.}},
\oauthor{\bsnm{Gan}, \binits{Q.}}:
Bayclump: Bayesian calibration and temperature reconstructions for clumped
  isotope thermometry.
Authorea Preprints
(2022)
\end{botherref}
\endbibitem

\bibitem[\protect\citeauthoryear{Petersen et~al.}{2019}]{petersen2019effects}
\begin{barticle}
\bauthor{\bsnm{Petersen}, \binits{S.V.}},
\bauthor{\bsnm{Defliese}, \binits{W.F.}},
\bauthor{\bsnm{Saenger}, \binits{C.}},
\bauthor{\bsnm{Da{\"e}ron}, \binits{M.}},
\bauthor{\bsnm{Huntington}, \binits{K.W.}},
\bauthor{\bsnm{John}, \binits{C.M.}},
\bauthor{\bsnm{Kelson}, \binits{J.R.}},
\bauthor{\bsnm{Bernasconi}, \binits{S.M.}},
\bauthor{\bsnm{Colman}, \binits{A.S.}},
\bauthor{\bsnm{Kluge}, \binits{T.}}, \betal:
\batitle{Effects of improved 17o correction on interlaboratory agreement in
  clumped isotope calibrations, estimates of mineral-specific offsets, and
  temperature dependence of acid digestion fractionation}.
\bjtitle{Geochemistry, Geophysics, Geosystems}
\bvolume{20}(\bissue{7}),
\bfpage{3495}--\blpage{3519}
(\byear{2019})
\end{barticle}
\endbibitem

\bibitem[\protect\citeauthoryear{Shah et~al.}{2020}]{shah2020research}
\begin{barticle}
\bauthor{\bsnm{Shah}, \binits{N.}},
\bauthor{\bsnm{Engineer}, \binits{S.}},
\bauthor{\bsnm{Bhagat}, \binits{N.}},
\bauthor{\bsnm{Chauhan}, \binits{H.}},
\bauthor{\bsnm{Shah}, \binits{M.}}:
\batitle{Research trends on the usage of machine learning and artificial
  intelligence in advertising}.
\bjtitle{Augmented Human Research}
\bvolume{5},
\bfpage{1}--\blpage{15}
(\byear{2020})
\end{barticle}
\endbibitem

\bibitem[\protect\citeauthoryear{Aggarwal
  et~al.}{2021}]{aggarwal2021diagnostic}
\begin{barticle}
\bauthor{\bsnm{Aggarwal}, \binits{R.}},
\bauthor{\bsnm{Sounderajah}, \binits{V.}},
\bauthor{\bsnm{Martin}, \binits{G.}},
\bauthor{\bsnm{Ting}, \binits{D.S.}},
\bauthor{\bsnm{Karthikesalingam}, \binits{A.}},
\bauthor{\bsnm{King}, \binits{D.}},
\bauthor{\bsnm{Ashrafian}, \binits{H.}},
\bauthor{\bsnm{Darzi}, \binits{A.}}:
\batitle{Diagnostic accuracy of deep learning in medical imaging: a systematic
  review and meta-analysis}.
\bjtitle{NPJ digital medicine}
\bvolume{4}(\bissue{1}),
\bfpage{65}
(\byear{2021})
\end{barticle}
\endbibitem

\bibitem[\protect\citeauthoryear{Lind and
  Pantigoso~Velasquez}{2019}]{lind2019performance}
\begin{botherref}
\oauthor{\bsnm{Lind}, \binits{E.}},
\oauthor{\bsnm{Pantigoso~Velasquez}, \binits{{\"A}.}}:
A Performance Comparison Between CPU And GPU In TensorFlow
(2019)
\end{botherref}
\endbibitem

\bibitem[\protect\citeauthoryear{Nocedal and
  Wright}{1999}]{nocedal1999numerical}
\begin{bbook}
\bauthor{\bsnm{Nocedal}, \binits{J.}},
\bauthor{\bsnm{Wright}, \binits{S.J.}}:
\bbtitle{Numerical Optimization}.
\bpublisher{Springer}, \blocation{???}
(\byear{1999})
\end{bbook}
\endbibitem

\end{thebibliography}

\end{document}